\documentclass[usenatbib,letterpaper]{mn2e}
\usepackage{graphicx}
\usepackage{ogonek}
\usepackage{amstext}

\title[Gravitational lensing in plasma]
  {Frequency-dependent effects of gravitational lensing within plasma}
\author[Adam Rogers]
 {Adam Rogers\thanks{E--mail: rogers@physics.umanitoba.ca} \\ Department of Physics and Astronomy, University of Manitoba, Winnipeg, MB, R3T 2N2, Canada}

\date{Accepted 2015 April 20. Received 2015 April 13; in original form 2015 February 16. \\ First published online 2015 May 22.}
\volume{451 (1)}
\pagerange{4536--4544} \pubyear{2015}

\begin{document}
\label{firstpage}
\maketitle

\begin{abstract}
The interaction between refraction from a distribution of inhomogeneous plasma and gravitational lensing introduces novel effects to the paths of light rays passing by a massive object. The plasma contributes additional terms to the equations of motion, and the resulting ray trajectories are frequency--dependent. Lensing phenomena and circular orbits are investigated for plasma density distributions $N \propto 1/r^h$ with $h \geq 0$ in the Schwarzschild space--time. For rays passing by the mass near the plasma frequency refractive effects can dominate, effectively turning the gravitational lens into a mirror. We obtain the turning points, circular orbit radii, and angular momentum for general $h$. Previous results have shown that light rays behave like massive particles with an effective mass given by the plasma frequency for a constant density $h=0$. We study the behaviour for general $h$ and show that when $h=2$ the plasma term acts like an additional contribution to the angular momentum of the passing ray. When $h=3$ the potential and radii of circular orbits are analogous to those found in studies of massless scalar fields on the Schwarzschild background. As a physically motivated example we study the pulse profiles of a compact object with antipodal hotspots sheathed in a dense plasma, which shows dramatic frequency--dependent shifts from the behaviour in vacuum. Finally, we consider the potential observability and applications of such frequency--dependent plasma effects in general relativity for several types of neutron star.
\end{abstract}

\begin{keywords}
gravitation -- plasmas -- stars: neutron -- pulsars: general
\end{keywords}

\section{Introduction}
\label{sec:intro}

It has been suggested that observations of black holes may be affected by general relativity (GR) and the frequency--dependent properties of plasma \citep{BKT09, plasmaLensingKerr} and that these effects may be relevant for black holes in X-ray binary systems \citep{BKT10}. Lensing effects modified by plasma surrounding galaxy scale lenses were also investigated by \citet{mao14}, requiring observations at MHz frequencies. In this work we will study ray--tracing under the influence of both gravitational lensing and refractive plasma effects following \citet{TBK13}, and examine frequency--dependent orbits for light rays. Deflection by both gravity and plasma was discussed by \citet{solarPlasmaDerivation} for the case of radio waves passing by the Sun, using the product of the gravitational and plasma refractive indices to derive a weak field approximation. \citet{solarPlasma} measured the deflection of the radio loud quasar 3C273 as it was occulted by the Sun and obtained an estimate of the electron density in the solar corona. In this work, we make use of the geometric optics approach developed by \citet{synge}, which has application to a diverse range of phenomena \citep{dressel, linet}. Fully covariant radiative transfer through a refractive medium was treated in depth by \citet{BB03}, and \citet{GR_plasma_2} discuss dispersion relations applicable to ray propagation in relativistic plasmas near black holes and rapidly rotating neutron stars (NSs). However, even the most basic dispersion effects and simple distributions of plasma near relativistic objects provide novel examples that differ from the vacuum behaviour of rays in the Schwarzschild space--time in interesting ways.

We review the theory and background of relativistic light bending including plasma effects in Section \ref{sec:theory}. In Section \ref{sec:orbits}, we study the trajectories of light rays through the plasma and consider waves near the plasma frequency with density varying as an inverse power law with index $h$. When $h=0$, the plasma frequency is constant and light rays behave like massive particles with mass given by the plasma frequency \citep{kulsrudLoeb}. When $h=2$, we find that the plasma term acts like an additive contribution to the angular momentum. The $h=3$ case has similarities with the behaviour of massless scalar fields on the Schwarzschild space--time. Section \ref{circularOrbits} includes a study of circular orbits for a general power-law index. We apply the analytical results to the pulse profiles of a compact object sheathed in plasma in Section \ref{sec:pulse}, and review relevant properties of compact objects in Section \ref{discussion}.

\section{Theory}
\label{sec:theory}
In this work we use natural units, with $G=c=1$, unless otherwise stated. As we proceed through the derivations of the main equations, we keep the $\hbar$ factors, but the numerical calculations are performed with $\hbar=1$. We follow the method and notation of \citet{synge}, and review the derivation of \citet{TBK13} to find the generalized trajectories of photons in the presence of an isotropic, inhomogeneous plasma distribution, though more general possibilities exist \citep{perlickGR}. Let us assume spherical symmetry and adopt the line element
\begin{equation}
\text{d}s^2=-A(r)\text{d}t^2+A(r)^{-1}\text{d}r^2+r^2\text{d}\theta^2+r^2\sin^2(\theta)\text{d}\phi^2
\end{equation}
with
\begin{equation}
A(r)= 1-\frac{r_{\text{g}}}{r}
\end{equation}
and $r_{\text{g}}=2M$ for the Schwarzschild metric. We write the coordinate indices as $(t,r,\theta,\phi)$, with the index $\alpha$ to sum over the spatial coordinates and $i,j,k$ to run over all four. Without loss of generality we consider orbits in the equatorial plane and set $\theta=\pi/2$. Now, suppose a distribution of plasma surrounds the mass $M$, which is located at the origin. The plasma has an index of refraction $n=n(x^{\alpha}, \omega)$ where the photon frequency is $\omega$, with effective energy $\hbar \omega= - p_i V^i$ with respect to the velocity of the plasma medium $V^i$. The linear momentum of the photon is $p^i=\hbar K^i$, where $K^i$ are the elements of the wave four-vector. The medium equation \citep{synge} gives the relation between the components of the photon four-momentum and the refractive index of the medium,
\begin{equation}
n^2=1+\frac{p_i p^i}{\left( p_j V^j \right)^2}.
\end{equation}
In the vacuum case $n=1$ and the standard condition for null geodesics $p_i p^i = 0$ is recovered. We describe photon paths through the plasma in the Schwarzschild space--time by the Hamiltonian in the geometric optics limit. Using the variational principle and the medium equation we have
\begin{equation}
H(x^i, p_i)=\frac{1}{2}\left[ g^{ij} p_i p_j + (n^2-1)\left( p_j V^j \right)^2 \right]=0.
\label{generalHamiltonian}
\end{equation}
Following \citet{TBK13}, we assume the plasma is a static, inhomogeneous medium such that $V^{t}=\sqrt{-g^{tt}}$, $V^{\alpha}=0$. This static condition simplifies the effective photon energy $\hbar \omega=-p_{t} \sqrt{-g^{tt}}$. With this condition there is no time or $\phi$ dependence of the Hamiltonian, so $p_{t}$ and $p_{\phi}$ are constant, and we restrict $p_{\theta}=0$ so that the orbit remains in the equatorial plane by symmetry. The time-like component $p_{t}$ is related to the vacuum photon energy
\begin{equation}
p_{t} = - \hbar \omega_{\infty}
\label{pt}
\end{equation}
and for a static medium this gives
\begin{equation}
p_{t} \sqrt{- g^{tt}}= - p^{t} \sqrt{ - g_{tt}}= - \hbar \omega( x^{\alpha})
\end{equation}
with the effective redshift relationship
\begin{equation}
\omega(r)=\frac{\omega_{\infty}}{A^{1/2}}
\label{omegaX}
\end{equation}
we have
\begin{equation}
p^{t}=\frac{\hbar \omega_{\infty}}{A}.
\label{p^t}
\end{equation}
The radial dependence of the photon frequency is due to the gravitational redshift supplied by the space--time metric, and we denote the photon frequency far from the origin $r \rightarrow \infty$ as $\omega_{\infty}$. For the sake of notational simplicity we will suppress the radial dependence when referring to the frequency variables.

In order to make analytical progress at this point we will introduce a specific form for the plasma frequency. Let us assume that the refractive index is of the general form
\begin{equation}
n^2=1- \frac{W^2}{\omega^2},
\label{nFreq}
\end{equation}
where $W = W( x^{\alpha},\omega)$. With equations (\ref{nFreq}) and (\ref{p^t}), the Hamiltonian in equation (\ref{generalHamiltonian}) simplifies such that
\begin{equation}
H(x^i,p_i)=\frac{1}{2}\left( g^{ij} p_i p_j + \hbar^2 W^2 \right) = 0.
\label{nullHamiltonian}
\end{equation}
The paths of light rays are then described in terms of the affine parameter $\lambda$ along the trajectory by
\begin{equation}
\frac{\text{d} x^i}{\text{d} \lambda} = \frac{\partial H}{\partial p_i} = g^{ij}p_j
\label{vel}
\end{equation}
and
\begin{equation}
\frac{\text{d} p_i}{\text{d} \lambda} = - \frac{\partial H}{\partial x^i} = -\frac{1}{2}g^{jk}_{,i} p_j p_k -\frac{\hbar^2}{2} (W^2)_{,i}.
\label{accel}
\end{equation}
These equations give the positions and momenta of photons through the space--time under the influence of gravity and the optical medium. The $r$ and $\phi$ components from equation (\ref{vel}) can be combined to eliminate the affine parameter $\lambda$, resulting in
\begin{equation}
\frac{\text{d} \phi}{\text{d} r} = \frac{p_{\phi}}{p_{r}} \frac{1}{r^2 A(r)}.
\end{equation}
The denominator can be simplified by using equation (\ref{nullHamiltonian}) to give $A p_{r}$ in terms of the constants $p_{t}$ and $p_{\phi}$, such that
\begin{equation}
A p_{r} = \pm p_{\phi} \sqrt{ \frac{p_{t}^2}{p_{\phi}^2} - A \left( \frac{1}{r^2}+ \frac{\hbar^2 W^2}{p_{\phi}^2} \right) }.
\label{pr}
\end{equation}
Here we take the positive solution corresponding to the case when both the $\phi$ and $r$ coordinates increase, and the negative solution with decreasing radial coordinate. To proceed, we write the constant $p_{\phi}$ in terms of the impact parameter $b$, such that
\begin{equation}
p_{\phi}=\hbar \omega_{\infty} b.
\label{pphi}
\end{equation}
Then we have
\begin{equation}
\frac{\text{d} \phi}{\text{d} r} = \pm \frac{1}{r^2 \left[ \frac{1}{b^2} \left( 1 - \frac{W^2}{\omega^2} \right) - \frac{A}{r^2} \right]^{1/2}}.
\label{dphidr}
\end{equation}
As expected, for $W \rightarrow 0$, equation (\ref{dphidr}) reduces to the vacuum case. This equation gives the paths of photons in the equatorial plane in the general case of strong deflection which is appropriate for orbits near the surface of the star. The refractive index of the plasma in this expression introduces frequency dependence to null geodesics \citep{noonan}. With the substitution $u=M/r$, outgoing rays have
\begin{equation}
\frac{\text{d} \phi}{\text{d} u} = - \frac{1}{\left[  \frac{M^2 n^2(u)}{b^2} - (1-2u)u^2 \right]^{1/2}},
\label{dphidu}
\end{equation}
where the potential dependence of $n$ on $u$ is explicitly stated. Now we consider an observer a great distance away at $r=D \rightarrow \infty$, such that the fiducial direction from the centre of the star to the observer corresponds to the impact parameter $b=0$. Since we are interested in describing rays which leave the surface of the star ($u_{R}=M/R$) and reach the observer ($u_{D}=0$), we define
\begin{equation}
\Delta = \int_0^{u_{R}}\frac{\text{d}u}{\left[ \frac{n^2(u)}{x^2} - (1-2u)u^2 \right]^{1/2}},
\label{Delta}
\end{equation}
with the additional simplification $x=b/M$, and taking the sign into account through the integration limits. This expression gives us the maximum viewing angle on the stellar surface that can be seen by the observer. In flat space $\Delta_{\text{max}}= \pi /2$ always, which means exactly half the surface of the star is visible. This is not the case in GR since gravitational lensing acts to redirect light rays such that a distant observer sees a larger apparent area and more of the stellar surface than would be possible in flat space. In fact, in extreme cases the observer can see portions of the surface multiple times (for $\Delta_{\text{max}} > \pi$).

It is also useful to describe the impact parameters of the rays which connect the surface of the star with the observer. Let us define $\delta$ as the angle between the radial and angular components of the photon momentum. At the surface of the star $r=R$, this is the angle with respect to the surface normal that the ray is emitted
\begin{equation}
\tan \delta = \frac{ \sqrt{ p^{\phi} p_{\phi}} }{ \sqrt{p^{r} p_{r}} }.
\end{equation}
Using equation (\ref{pr}) leads to
\begin{equation}
b=\frac{R n(R)}{ A(R)^{1/2} } \sin \delta
\label{b}
\end{equation}
where the maximum impact parameter $b_{\text{max}}$ occurs for $\delta=\pi/2$. This formula reduces to the standard expression for the emission angle in vacuum when $n=1$ \citep{pfc,bel02}. The presence of the refractive index within the impact parameter expression predicts that the portion of the stellar surface visible to a distant observer shrinks as $n \rightarrow 0$.

\begin{figure}
\centering
\includegraphics[bb= 199 223 423 566, clip, scale=0.8]{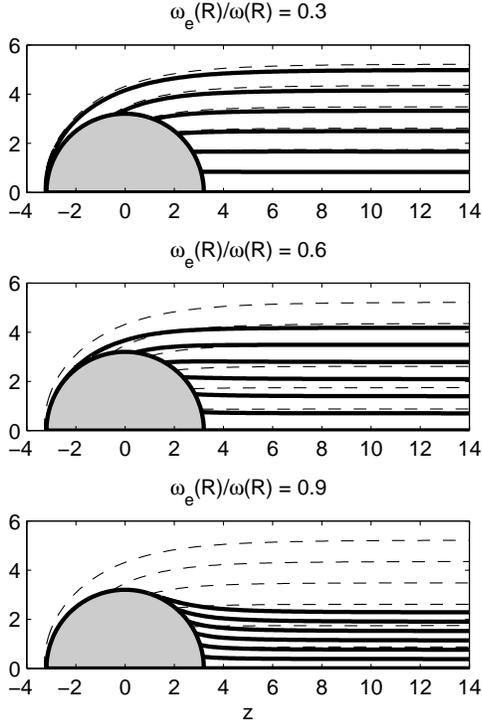}
\caption{Examples of ray--tracing from the surface of a compact object for a sample of photon frequencies, denoted by surface values $\omega_{\text{e}}(R)/\omega(R)$. The vacuum solution is shown on all panels as a set of thin dashed lines. Impact parameters between $0$ and $b_{\text{max}}$ are used to illustrate the shrinking of the visible surface for rays that approach the plasma frequency. This example makes use of a highly relativistic star with $R/r_{\text{g}}=1.60$, and a plasma frequency of the form $\omega_{\text{e}}^2=1/r^3$.}
\label{fig:rayTrace}
\end{figure}

Specific numerical examples require a choice of $W(x^{\alpha}, \omega)$. The simplest option is to pick
\begin{equation}
W^2=\omega_{\text{e}}^2=\frac{4 \pi e^2 N(r)}{m}
\label{plasmaFreqDef}
\end{equation}
where $\omega_e(r)$ is the plasma frequency, $e$ and $m$ the electron charge and mass respectively, and $N(r)$ the plasma number density. Together with equations (\ref{nFreq}) and (\ref{generalHamiltonian}), equation (\ref{plasmaFreqDef}) gives the dispersion relation for rays in a cold, non-magnetized plasma. We will make use of this form in the following sections, but we stress that our results to this point are general for $n=n(x^{\alpha},\omega)$. Propagation through this medium requires that the photon has $\omega(r) > \omega_{\text{e}}(r)$ everywhere. In the case when $\omega \gg \omega_{\text{e}}$ or the plasma density vanishes $N(r)=0$, we recover the vacuum case $n=1$. The condition for rays to propagate from the surface to infinity is
\begin{equation}
\omega_{\infty} > \omega_{\text{e}}(R) A(R)^{1/2}.
\label{propCond}
\end{equation}
Consider the situation illustrated in Fig. \ref{fig:rayTrace}. In this figure we show the behaviour of light rays with a range of asymptotic frequencies $\omega_{\infty}$. We use the surface value $\omega_{\text{e}}(R) / \omega(R)$, with equations (\ref{omegaX}) and the propagation condition (\ref{propCond}) to label the frequency ratio. Tracing rays from the stellar surface numerically requires a specific plasma density distribution, so we choose a physically motivated example based on the Goldreich--Julian (GJ) density \citep{JG69}. Normal pulsar magnetospheres contain plasma with densities that are sustained by GJ currents, transferring charge carriers from the surface of the NS to the magnetosphere. These currents therefore provide a lower limit for the density of charge carriers in a standard magnetosphere. Using the plasma frequency (equation \ref{plasmaFreqDef}), let us consider a radial power-law density
\begin{equation}
N(r)=\frac{N_0}{r^h},
\label{powerLawDensity}
\end{equation}
where $h \geq 0$, such that
\begin{equation}
\omega_{\text{e}}^2=\frac{k}{r^{h}}.
\label{omegaN}
\end{equation}
Since the GJ density depends on the strength of the polar magnetic field we use $h=3$ giving a $1/r^3$ form. Let us consider a highly relativistic star with $R/r_{\text{g}}=1.60$, and assume a plasma frequency $\omega_{\text{e}}$ with $k=1$ as an arbitrary constant. We plot the rays with positive impact parameter ranging from $0$ to $b_{\text{max}}$ as thick dotted lines. All frequencies satisfy the condition for propagation given in equation (\ref{propCond}).

Classical turning points exist for the ray paths where $\text{d}r/\text{d}\phi=0$. For all rays with impact parameter $b < b_{\text{max}}$ these turning points are within the spherical surface of the star and have no effect on the ray paths. For rays with the maximum impact parameter the turning point is exactly on the surface of the star, indicating a ray that just grazes the stellar surface. The rays which have larger impact parameters than $b_{\text{max}}$ do not connect the stellar surface to a distant observer.

\section{Null orbits within refractive media}
\label{sec:orbits}

In this section, we are interested in the gravitational lensing effect on rays passing by the star. These rays approach from infinity and have impact parameters $b>b_{\text{max}}$. The turning points for these trajectories, $r_{\text{p}}$, are found when the denominator of equation (\ref{Delta}) vanishes:
\begin{equation}
1-\left( 1-\frac{2M}{r}\right)\left( \frac{b^2}{r^2}+\frac{\omega_{\text{e}}^2}{\omega_{\infty}^2} \right) = 0.
\end{equation}
Using a power-law density distribution gives rise to a polynomial with order that depends on the power-law exponent
\begin{equation}
r^{h+1} - b^2 r^{h-1} + 2M b^2 r^{h-2} - \frac{k}{\omega_{\infty}^2} r  + 2M\frac{k}{\omega_{\infty}^2} = 0.
\label{turningPts}
\end{equation}
The solutions of this equation signify the boundary between regions in which $\Delta$ is real and complex valued. A trajectory that reaches a turning point has vanishing radial momentum and subsequently reverses direction. When the effect of refraction is significant, even light rays near the line of sight with small impact parameters can be strongly affected by the interaction with the medium and turned away. In this case the plasma behaves like a scattering medium, and it is only far from the origin that the vacuum-type lensing behaviour is recovered.

We define the energy and effective potential in analogy with the vacuum case by associating the $p_{\phi}$ with the angular momentum of a photon, and the time component $p_{t}$ with the vacuum energy of a photon. We call the radial rate of change with respect to the curve parameter $\dot{r}$, and equations (\ref{vel}) and (\ref{pr}) can be used to write
\begin{equation}
\dot{r}^2=p_{t}^2-\left( 1-\frac{2M}{r}\right) \left( \frac{L^2}{r^2} + \frac{k}{r^h} \right) = p_{t}^2-V_{\text{eff}}.
\label{nrg}
\end{equation}
The effective potential of the Schwarzschild space--time with a plasma distribution is
\begin{equation}
V_{\text{eff}} = \left( 1-\frac{2M}{r} \right)\left( \frac{L^2}{r^2} + \frac{k}{r^h} \right)
\label{effPot}
\end{equation}
where $V_{\text{eff}}=V + V_{\text{p}}$, with $V$ and $V_{\text{p}}$ the vacuum and plasma contributions, respectively. In the absence of plasma $V_{\text{p}}=0$ ($k=0$) the potential reduces to the vacuum case for null geodesics.

\begin{figure}
\centering
\includegraphics[scale=0.65, bb= 110 222 478 565, clip=true]{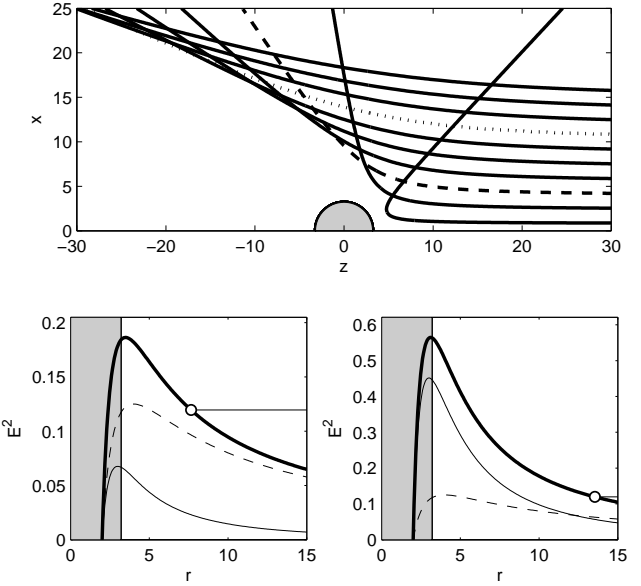}
\caption{Ray--tracing and effective potential, $h=1$ case: an example of ray--tracing in the case of $\omega^2_{\text{e}}=k/r$ for rays that have $\omega_{\text{e}}(R) / \omega(R) = 0.99$, for $R=3.2$ and $M=1$. The potential acting on the dashed ray is shown in the lower--left panel and the potential acting on the dotted ray in the lower right. The effective potential is the thick line, the vacuum and plasma contributions are the thin and dashed lines respectively.}
\label{rayTraceh1}
\end{figure}

\begin{figure}
\centering
\includegraphics[scale=0.66, bb= 110 222 478 518, clip=true]{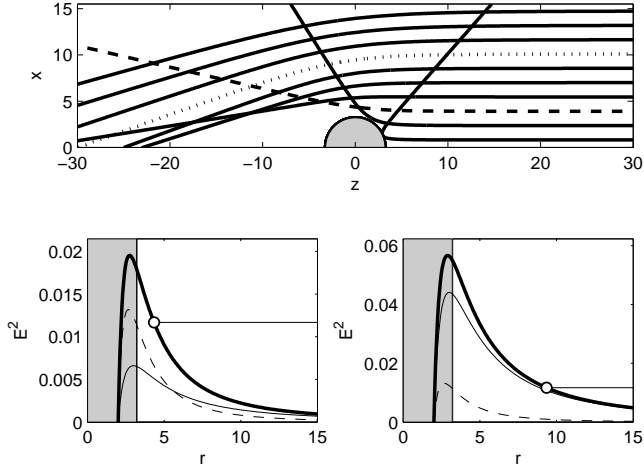}
\caption{Ray--tracing and effective potential, $h=3$ case: ray--tracing with $\omega^2_{\text{e}}=k/r^3$ for rays near the plasma frequency $\omega_{\text{e}}(R) / \omega(R) = 0.99$, and other parameters as in Fig. \ref{rayTraceh1}. Note that no trajectory actually touches the surface of the object, and that the closest approach in this figure is slightly exaggerated by the thickness of the line.}
\label{rayTrace}
\end{figure}

For the $h=0$ case $\omega_{\text{e}}$ is constant, and photons behave like massive particles with effective mass equal to the plasma frequency, a scenario that was explored by \citet{kulsrudLoeb}, \citet{BB03} and \citet{TBK13}. These photon paths are identical to the well-studied time-like paths in the vacuum Schwarzschild case. However, it is far more interesting to explore the behaviour for other, less trivial plasma density distributions. The $h=1$ case is plotted in Fig. \ref{rayTraceh1} for the trajectories of $10$ rays past the mass $M=1$ with radius $R=3.2$, and shows the corresponding effective potential for two of these rays. We plot the square of the photon effective energy $E^2 = p_{t}^2$, the effective potential $V_{\text{eff}}$, and the contributions from the vacuum and plasma parts $V$ and $V_{\text{p}}$. The effect of the plasma dominates over gravity for rays near the plasma frequency, and this scattering behaviour is more pronounced at lower $h$. For the $h=3$ case, shown in Fig. \ref{rayTrace}, the plasma density drops off quickly enough that we recover behaviour similar to lensing in vacuum for rays further from the compact object even at low ratios of photon to plasma frequency. In the figures we arbitrarily set $k=1$ for these numerical examples.

The $h=2$ case has the plasma term act like an additional contribution to the angular momentum. To make this connection explicit, consider the turning points from equation (\ref{turningPts}) with the angular momentum $L=\omega_{\infty} b$:
\begin{equation}
r^3 -  \frac{(L^2+k)}{\omega_{\infty}^2} r + \frac{(L^2+k)}{\omega_{\infty}^2} r_{\text{g}} = 0
\end{equation}
the solutions are known for this specific form of third order polynomial
\begin{eqnarray}
r_{\text{m}} &=& -\frac{2}{\omega_{\infty}} \sqrt{\frac{L^2+k}{3}} \nonumber \\
      \label{rmcos}
      & & \times \cos \left[ \frac{1}{3} \cos^{-1} \left( \frac{3\sqrt{3}}{2} \frac{\omega_{\infty} r_{\text{g}}}{\sqrt{L^2+k}}\right) +\frac{2\pi m }{3}\right]
\end{eqnarray}
with $m=-1,0,+1$, analogous to the vacuum case with $L^2 \rightarrow L^2+k$.

\subsection{Circular orbits}
\label{circularOrbits}
Circular orbits with radii $r_{\text{c}}$ occur at extrema of the potential, where the first derivative vanishes. For general $h$ this is
\begin{equation}
\frac{\text{d}V_{eff}}{\text{d}r}=\frac{-2L^2}{r^3}\left( 1-\frac{3M}{r} \right)-\frac{k}{r^{h+1}}\left[ h-\frac{2M}{r}(h+1) \right].
\label{Vdiff}
\end{equation}
Stability is determined from the sign of the second derivative
\begin{equation}
\frac{\text{d}^2V_{\text{eff}}}{\text{d}r^2}=\frac{6L^2}{r^4}\left(1-\frac{4M}{r} \right) + \frac{k(h+1)}{r^{h+2}}\left[ h-\frac{2M}{r}(h+2) \right].
\end{equation}
We summarize the behaviour of the circular paths in Table \ref{table1} for integer $h=0,...,3$. It is also easy to find the corresponding angular momentum necessary for circular motion in general. Setting equation (\ref{Vdiff}) equal to zero and solving for the angular momentum gives
\begin{equation}
L^2=\frac{M r_{\text{c}}^2}{(r_{\text{c}}-3M)} \frac{k}{r_{\text{c}}^{h-1}} \left[ \frac{h+1}{r_{\text{c}}} - \frac{h}{2 M} \right].
\end{equation}
This is the angular momentum that is required to keep a photon in a circular path around a compact object surrounded by a plasma distribution with a power-law density. The radii of the circular orbits is listed in Table \ref{table1} as $r_{\text{c}}$, along with the corresponding density index $h$. Only the $h=0$ case admits both solutions, which are physical and become stable when they coincide at $r=6M$. For the other cases, only one of the roots is positive. All of the solutions have $L \rightarrow \infty$ as $r \rightarrow 3M$.

\begin{table}
\caption{Circular orbit radii for corresponding integer power-law density index $h$. The radial range where unstable circular orbits exist is also given.}
\scalebox{0.83}{
\noindent%
\begin{tabular}{|c|c|c|}
\hline
Index & Radius of orbit & Range\\ \hline
$h=0$ & $r_{\text{c}}= \frac{L^2}{2Mk}\left( 1 \pm \sqrt{1-\frac{12M^2k}{L^2}} \right)$ & $3M-6M$\\
\hline
$h=1$ & $r_{\text{c}}=\frac{2M}{k}\left( k - \frac{L^2}{2M} + \sqrt{ k^2 + \frac{L^4}{4M^2} + \frac{1}{2}\frac{kL^2}{M} } \right)$ & $3M-4M$\\
\hline
$h=2$ & $r_{\text{c}}=3M$ & -- \\
\hline
$h=3$ &
$r_{\text{c}}=\frac{3M}{2L^2}\left( L^2-\frac{k}{2M} + \sqrt{ L^4 + \frac{k^2}{4M^2} + \frac{7}{9}\frac{kL^2}{M} } \right)$ & $\frac{8M}{3}-3M$ \\
\hline
\end{tabular}
\label{table1}}
\end{table}

Introducing terms into the effective potential through refractive plasma brings new behaviour to the trajectories of light rays, some of which correspond analogously to known cases. For example, the $h=3$ case is related to the effective potential found in the Klein--Gordon equation that describes the evolution of scalar field perturbations on the Schwarzschild space--time \citep{chandrasekharDetweiller1975, natarioSchiappa04, skakalaVisser10}. To proceed, we assume the energy density of the scalar field to be small and its influence on the space--time geometry negligible. Due to spherical symmetry, the radial and angular parts of the Klein--Gordon equation can be separated \citep{zerelli70}, and the field written in terms of vector and spherical harmonics. This radial part is the Regge--Wheeler equation \citep{reggeWheeler57} for massless, electromagnetic ($s=1$) and gravitational ($s=2$) perturbations on the Schwarzschild background
\begin{equation}
\frac{\text{d}\psi_l^2}{\text{d}r_*^2}+\left(\omega^2-V_{\text{RW}} \right) \psi_l^s=0
\end{equation}
where radial coordinates with an asterisk denote the use of a `tortoise coordinate',
\begin{equation}
\frac{\text{d}r_*}{\text{d}r}=\left( 1-\frac{2M}{r} \right)^{-1}.
\end{equation}
The potential governing the behaviour of the perturbation is
\begin{equation}
V_{\text{RW}}=\left( 1-\frac{2M}{r} \right) \left[ \frac{l(l+1)}{r^2}+\frac{2M(1-s^2)}{r^3}\right],
\end{equation}
with the orbital quantum number $l$ and spin given by $s$, along with the condition $l \geq s$ \citep{LunFackerell74}. Writing $\Lambda=l(l+1)$ and $\beta=1-s^2$, the solution for circular orbits is then
\begin{equation}
r_{\text{RW}}=\frac{3M}{2\Lambda}\left( \Lambda-\beta + \sqrt{\Lambda^2+\beta^2+\frac{14}{9}\beta\Lambda} \right).
\label{rsh3}
\end{equation}
Now consider the case of a plasma surrounding a compact object with a density profile that drops off as $1/r^3$, such that $h=3$ in equation (\ref{effPot}). The resulting effective potential is a direct analogy to the potential that arises in the Regge-Wheeler equation. The solutions for circular orbits $r_{\text{RW}}$ and $r_{\text{c}}$ for $h=3$ in Table \ref{table1} can be transformed into one another given the substitutions $L^2 \rightarrow l(l+1)$ and $k \rightarrow 2M(1-s^2)$. The analogy works directly for $s=0$, and reproduces the vacuum case for $s=1$ ($k=0$) as expected for photons. For $s=2$ we must allow $k<0$ which corresponds to a refractive index that has $n>1$ everywhere, but is an unphysical density distribution for plasma. However, with this caveat we are able to reproduce the potential and circular orbits of massless scalar fields of arbitrary spin. While we work in the geometric optics limit, our trajectories point along the propagation directions of the corresponding solutions to the Regge-Wheeler equation. Thus, the interaction of passing electromagnetic radiation with a gravitational field and a plasma that has a density distribution that drops off as $1/r^3$ is in some sense analogous to the more general interaction of a massless boson and scalar field perturbation on the Schwarzschild background.

\section{Effects on pulse profiles and timing}
\label{sec:pulse}
As a physically motivated scenario, let us consider an NS embedded within a plasma with bright polar emission regions near the surface that produce observable broad-band pulsations. This toy model provides an example of the significance refraction can have on the self-lensing of a compact object. The basic model and viewing geometry used here were thoroughly discussed by \citet{pfc}, and the interested reader should consult that work for full details.

Let us define the observer's coordinate system with respect to a plane perpendicular to the line of sight (the `detector') which collects the rays from a distant object. The detector surface records the flux received from the object (the `image') over the angles $\theta'$ and $\phi'$ such that an element of solid angle in the observer coordinate system is given by
\begin{equation}
\text{d} \Omega'=\sin(\theta') \text{d} \theta' \text{d} \phi' \approx \theta' \text{d} \theta' \text{d} \phi'
\end{equation}
under the appropriate small angle assumption for $\theta'$. This angular patch is related to the impact parameter in a simple way by $b=D \theta'$ which gives
\begin{equation}
\text{d}\Omega'=\frac{1}{D^2}b \text{d}b \text{d} \phi'
\end{equation}
where $b$ is given by equation (\ref{b}) and varies from $0$ to $b_{\text{max}}$. Since the star is spherically symmetric $\phi'=\phi$. Thus we can imagine tracing each patch of image back to some location on the surface of the star given by the angles $\Delta$ (equation \ref{Delta}) and $\phi$. In this sense $\Delta$ translates the two dimensional solid angle element of the observer to a corresponding surface area element on the star. To generate pulse profiles we must find the flux from a bright emission region at the pole of the star and describe the projection of this feature on the observer's sky.

The line of sight passes through the origin of the detector plane and the centre of the star, taken to be the origin of a spherical coordinate system fixed relative to the surface of the star (the `object' coordinates). Let a circular emission region (a `hotspot') be centred on the polar cap with an opening angle $\theta_{\text{c}}$, and the centre of the polar cap be denoted by a unit vector $\mbox{\boldmath$\hat{C}$}$. Let us also define a vector along the line of sight, $\mbox{\boldmath$\hat{L}$}$, such that the angle between $\mbox{\boldmath$\hat{L}$}$ and the cap centre $\mbox{\boldmath$\hat{C}$}$ is called $\theta_0$.

In general $\theta_0$ is a time--dependent quantity since the star rotates. To express this as a function of time, we define the phase $\gamma_{\text{P}}=\Omega t+\gamma_0$ where $\Omega=2\pi /P$ with $P$ the rotation period and $\gamma_0$ an arbitrary phase constant. Then the angle between the rotation axis $\mbox{\boldmath$\hat{r}$}$ and line of sight is $\chi=\cos^{-1} \left( \mbox{\boldmath$\hat{r}$} \cdot \mbox{\boldmath$\hat{L}$} \right)$ and the angle between the rotation axis and the centre of the hotspot is $\xi = \cos ^{-1} \left( \mbox{\boldmath$\hat{r}$} \cdot \mbox{\boldmath$\hat{C}$} \right)$. The orientation of the cap can then be expressed with respect to these angles as
\begin{equation}
\theta_0 (t) = \cos ^{-1} \left[ \cos\chi \cos \xi - \sin \chi \sin \xi \cos \gamma_P (t) \right]
\label{theta0Time}
\end{equation}
where the time dependence of $\theta_0$ has been explicitly stated.

In the object coordinates, an arbitrary point on the cap boundary is given by ($\theta_{\text{c}}$, $\Phi$), and the corresponding point on the observer's detector is ($\theta$, $\phi_{\text{b}}$). In general the two coordinate systems are rotated about the $\mbox{\boldmath$\hat{L}$}$-$\mbox{\boldmath$\hat{C}$}$ plane. Using the transformation between the cartesian components of the two systems, we arrive at an expression for $\phi_{\text{b}}(\theta)$ in terms of the observer coordinates,
\begin{equation}
\phi_{\text{b}}(\theta)=\cos^{-1}\left( \frac{\cos \theta_{\text{c}} - \cos \theta_0 \cos \theta}{\sin \theta_0 \sin \theta} \right)
\label{phib}
\end{equation}
for a given $\theta$. Thus we are able to determine the boundary of the hotspot in a slice along $\theta$ in the observer coordinates for a given orientation of the star (i.e., a given $\theta_0$) and a given cap size $\theta_{\text{c}}$. The hotspot extends over the range $-\phi_{\text{b}}$ to $\phi_{\text{b}}$, and we denote this range as $H(\theta;\theta_0,\theta_{\text{c}})=2\phi_{\text{b}}$, simplifying the observer's solid angle element. Determining the observable flux from a hotspot therefore only requires a one--dimensional integral over the impact parameter $b$.

In general the definition of $\phi_{\text{b}}$ is more complicated than described above. In fact, $\phi_{\text{b}}$ is not defined for all arbitrary orientations of the cap since equation (\ref{phib}) requires $\theta_0 + \theta_{\text{c}} \leq \theta \leq \pi$ and $\theta_0-\theta_{\text{c}} \geq 0$ for example. However, the conditions necessary to continuously extend $\phi_{\text{b}}$ are given in a variety of sources including \citet{pfc} and \citet{dabrowski} who include a thorough discussion for cases when multiple imaging is important. To compare our results with the vacuum case we introduce one final substitution. In equation (\ref{Delta}) we set $x=b/M$, which brings us to
\begin{equation}
\text{d}\Omega' = \frac{M^2}{D^2} H(\theta[x]; \theta_0, \theta_{\text{c}} ) x \text{d}x.
\end{equation}

\begin{figure}
\centering
\includegraphics[scale=0.65]{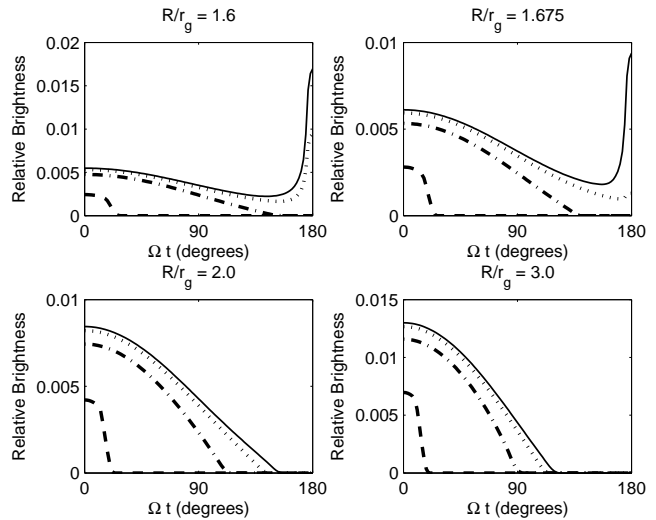}
\caption{Pulse profiles of a single hotspot, $h=3$. The pulses are plotted for an orthogonal configuration with $\chi=\xi=\pi/2$, and frequency ratios $\omega_{\text{e}}(R) / \omega(R)=0.99$ (dashed line), $\omega_{\text{e}}(R) / \omega(R) = 0.60$ (dashed-dotted), $\omega_{\text{e}}(R) / \omega(R) = 0.30$ (dotted). The aperture angle is $\theta_{\text{c}}=5^{\circ}$. The vacuum case is plotted as a thin, solid line. The upper left panel has $R/r_{\text{g}}=1.6$, the upper right $R/r_{\text{g}}=1.675$, lower left $R/r_{\text{g}}=2.0$ and lower right $R/r_{\text{g}}=3.0$. The pulse profiles narrow as $\omega(R)$ approaches $\omega_{\text{e}}(R)$.}
\label{pulseSingle}
\end{figure}

\begin{figure}
\centering
\includegraphics[scale=0.65]{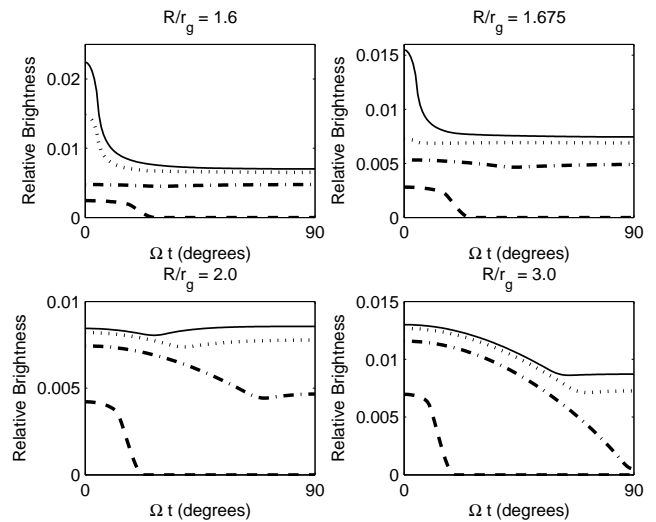}
\caption{Pulse profiles of antipodal hotspots, $h=3$. The pulse profiles for two antipodal hotspots are plotted. Parameters are identical to those used in Fig. \ref{pulseSingle}.}
\label{pulseDouble}
\end{figure}

To find the flux of a surface bright spot we must use the specific intensity. \citet{bicakHadrava} have shown that the generalization of the invariant relationship between specific intensity and frequency along a ray in a refractive medium is
\begin{equation}
\frac{I_{\text{obs}}}{\omega_{\text{obs}}^3 n_{\text{obs}}^2}=\frac{I_{\text{em}}}{\omega_{\text{em}}^3 n_{\text{em}}^2}=\text{constant}
\label{IobsRes}
\end{equation}
where the labels obs and em denote the observed and emitted quantities, respectively. Taking the observation point to be infinitely far from the star, $\omega_{\text{obs}}=\omega_{\infty}$, $n_{\text{obs}}=1$, and assuming the emission point on the surface $\omega_{\text{em}}=\omega_{\infty}A(R)^{-1/2}$ and $n_{\text{em}}=n(R)$. By equating the observed and emitted intensity through this relationship we obtain
\begin{equation}
I_{\text{obs}} =  \frac{A^{3/2}}{n^2(R)} I_{\text{em}}.
\end{equation}
For simplicity we take $ I_{\text{em}} = n^2(R) I_0 f_{\text{B}}(\delta) $ where $ I_0 $ is a constant and $ f_{\text{B}} $ a function of emission angle to accommodate anisotropic emission effects
\begin{equation}
I_{\times{obs}}=\left( 1- \frac{2M}{R} \right)^{3/2} I_0 f_{\text{B}}(\delta).
\label{Iobs}
\end{equation}
Other choices for $I_{\text{obs}}$ are also possible, such as approximating the specific intensity via the brightness temperature \citep{lorimerKramer2005}. The emission angle at the surface is a function of $x$ since
\begin{equation}
\delta=\sin^{-1}\left( x/x_{\text{max}} \right)
\end{equation}
where $x_{\text{max}}=b_{\text{max}}/M$.
The flux at a given frequency $\omega$ is
\begin{eqnarray}
F_{\omega} &=& \left(1 -\frac{2M}{R}\right)^{3/2} \frac{M^2}{D^2} \nonumber \\
    \label{Flux}
    & & \times \int_0^{ x_{\text{max}} } I_0 f_{\text{B}}(\delta[x]) H(\theta[x]; \theta_0, \theta_{\text{c}} ) x \text{d}x.
\end{eqnarray}
The result looks similar to that obtained by \citet{pfc}; however, the key differences come from $\Delta$ and the range of $x$, since it depends on the maximum impact parameter which is now dependent on frequency. The presence of an antipodal hotspot simply requires a second component with $\theta_0 \rightarrow \pi-\theta_0$. To generate light curves we recall the time dependency of $\theta_0$ and plot $F_{\omega}$ for a variety of frequency ratios $\omega_{\text{e}}(R) / \omega(R)$ and mass-radius relationships $R/r_{\text{g}}$. The ratios we used start with a case that demonstrates multiple imaging of the stellar surface. We decrease this ratio in each subsequent plot to show how the pulse profiles change for larger ratios of radius to mass. We consider a constant size polar cap fixed with aperture $\theta_{\text{c}}=5^{\circ}$ in an orthogonal configuration with $\chi=\xi=\pi$. We plot the $h=3$ case in Fig. \ref{pulseSingle}, and an analogous example with two antipodal hotspots in Fig. \ref{pulseDouble}. We emphasize the change in morphology of the pulse profiles as a function of frequency by assuming isotropic emission, setting $f_{\text{B}}(\delta)=1$. The calculations are done in terms of the ratio $\omega_{\text{e}}(R) / \omega(R)$, so the specific value of $k$ will affect only the frequency at which plasma effects become significant, and will leave the morphology of the calculated pulse profiles unchanged.

\begin{figure}
\centering
\includegraphics[scale=0.65]{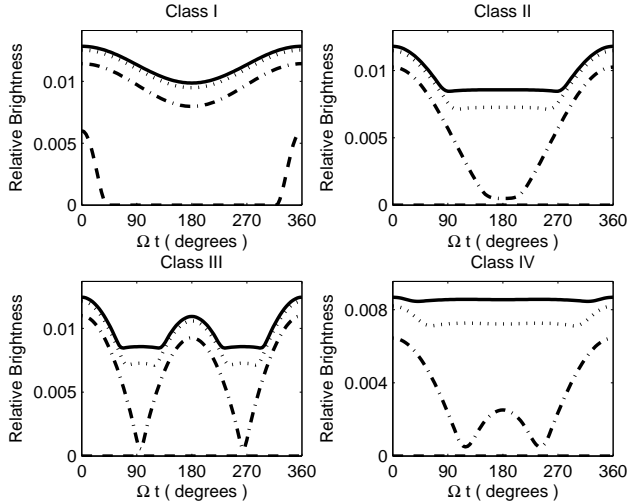}
\caption{A demonstration of pulse classes for a NS with $R/r_{\text{g}}=3$. All other parameters are as in Fig. \ref{pulseSingle}. We plot plasma frequency ratios $\omega_{\text{e}}(R) / \omega(R)=0.99$ (dashed line), $\omega_{\text{e}}(R) / \omega(R) = 0.60$ (dashed-dotted), $\omega_{\text{e}}(R) / \omega(R) = 0.30$ (dotted), which is similar to the vacuum profile for all classes. The upper--left panel has $\chi=20^{\circ}$, $\xi=30^{\circ}$, the upper--right panel $\chi=30^{\circ}$, $\xi=60^{\circ}$, the lower--left panel has $\chi=60^{\circ}$, $\xi=80^{\circ}$ and the lower--right panel has $\chi=20^{\circ}$, $\xi=80^{\circ}$. Except in class I, the profiles completely vanish nearest the plasma frequency.}
\label{pulseClasses}
\end{figure}

We vary the angles $\chi$ and $\xi$ in Fig. \ref{pulseClasses} to demonstrate the frequency--dependent pulse morphologies for a variety of viewing angles and system geometries. With the convenient classification scheme of \citet{bel02} we show separate examples of each pulse class for antipodal configurations. In this scheme, class I has a primary spot which is always visible, and an antipodal spot which is never seen. In class II profiles, the primary spot is always seen and the antipodal spot is only visible for part of the time. Class III profiles are generated when the primary and antipodal spots are both visible for fractions of the time and class IV has both spots seen at any time, producing a relatively constant profile. We plot the profiles for the $h=3$ case which shows how the profiles evolve as a function of $\omega$.

Another significant effect for pulse profiles is the time delay, which is also affected by the medium. Consider the arrival times of photons leaving the stellar surface and travelling to a distant observer
\begin{equation}
t(R)=\int_{t(R)}^{t(\infty)}dt=\int_{R}^{\infty}\frac{dt}{d\lambda}\frac{d \lambda}{dr}dr.
\end{equation}
The time delay $\Delta t(R)$ for a given impact parameter is found by comparing to the fiducial value $b=0$. For photons emitted from the surface at $R$, we have
\begin{equation}
\Delta t(R) = \int_{R}^{\infty} \frac{1}{A}\left[ \frac{1}{ \left( n^2-\frac{Ab^2}{r^2} \right)^{1/2} } - \frac{1}{n} \right] dr
\end{equation}
which reduces to the vacuum case when $n=1$.

\section{Discussion}
\label{discussion}

The diverse population of NSs is defined by a wide range of magnetic fields, varying from the rapidly spinning millisecond pulsars (MSPs) with fields from $10^{8}$ to $10^{10}$ G \citep{MSPB} to the slowly rotating magnetars which have super--critical QED scale fields between $10^{14}$ and $10^{15}$ G \citep{magnetar1}. Generally, NSs possess complex magnetospheres containing plasma that can have a significant effect on pulse profiles \citep{plasmaMagnetosphere, aronsBarnard, barnardArons, WWH14}, though the exact mechanism responsible for radio emission in the magnetosphere is not well understood \citep{hankinsRadio}. The GJ number density provides a lower limit for the charge density in the neighbourhood around an NS, and is given by
\begin{equation}
N_{\text{GJ}} = -\frac{ \Omega B }{2 \pi e c} \cos \xi \approx 7 \times 10^{-2} \frac{B_{\text{z}}}{P}
\label{GJdensity}
\end{equation}
in units of particles per cm$^3$, with $P$ the rotation period in seconds and $B_{\text{z}}$ (G) the field component along the rotation axis in the approximation above \citep{JG69}. Since $N_{\text{GJ}}$ is a lower limit we use the proportionality $\kappa$ to discuss the magnetospheric density $N_0=\kappa N_{\text{GJ}}$ \citep{GJpropto}. This expression gives a physical motivation for the $h=3$ plasma density index discussed throughout this work. The ray--tracing discussed in Section \ref{sec:theory} can be used for rays passing through a pulsar wind, introducing relativistic and plasma effects into models of binary eclipses for a double NS or NS--black hole binary that is nearly edge on with respect to the observer. In the case of an aligned rotator we expect a wind density proportional to $1/r$, such that $h=1$ \citep{wind1, magWind1, wind2, magWind2}, shown in Fig. \ref{rayTraceh1}. It is plausible that both lensing and plasma may affect the eclipse mechanism for as-yet undiscovered compact binary systems with an edge on orientation \citep{Keane, SKA}.

Normal radio pulsars show a rich variety of frequency--dependent behaviours \citep{150_PSRs_in_radio}. A proportionality constant $10^3 < \kappa < 10^5$ is needed to describe the dense environment surrounding these objects \citep{plasmaFreq1}. Generally the simplification of cold plasma is unrealistic since matter within the light cylinder is strongly magnetized in corotation with the NS and highly relativistic. Particles moving along the field lines have a Lorentz factor $\gamma=E / m_{\text{e}} c^2$, giving an effective plasma frequency
\begin{equation}
\omega_{\text{eff}} \approx \frac{\omega_{\text{e}}}{ \left< \gamma \right>^{1/2} },
\end{equation}
where the right-hand side is summed and averaged over the positron and electron distribution functions \citep{gammaFactor, cutoffFrequency}. The relativistic case gives a much lower cutoff frequency, well below $100$ MHz \citep{cutoffFrequency, radioCutoff3}. Thus, despite the low frequencies involved the interplay between GR and plasma effects in pulsars may potentially be within observable ranges \citep{radioLow2, radioLow1, LOFAR}. It is also significant to note the existence of a subset of radio pulsars with pulse widths that broaden with increasing frequency \citep{150_PSRs_in_radio}, also seen in Figs. \ref{pulseSingle} and \ref{pulseDouble} for the toy light curves generated here.

Due to their fast rotation rate, the radio emission region is thought to be closest to the NS surface in MSPs \citep{alparMSPEvolution, MSPEvolution}. If a low-frequency component of the radio emission from MSPs does originate sufficiently near the surface \citep[ie, $r/R<10$;][]{gil_kijak_1993, kijak_gil_1998}, it may be plausibly affected by both relativistic and plasma effects. The significance of rotation for a quickly spinning MSP, especially due to induced oblateness of the surface, becomes increasingly important for the pulse profiles and time delays, making the Schwarzschild case a first approximation \citep{cadeau07}.

At the opposite extreme from MSPs, anomalous X-ray pulsars (AXPs) and soft gamma repeaters (SGRs) appear to be NSs with magnetic fields above the QED critical field $B_{\text{QED}} = 4.4 \times 10^{13}$ G \citep{mReview}, most commonly described by the magnetar model \citep{magnetar1, magnetar3, magnetar2}. The lack of persistent radio emission above $500$ MHz \citep{istomin, magnetarRadioLow1} and the transient, highly-variable radio emission observed from a subset of these objects \citep{1E1547radio, malov14} suggests that the density of plasma in the magnetospheres of magnetars is lower than in normal radio pulsars \citep{istomin, malov14}. While gravitational effects are expected to dominate photon trajectories outside of magnetars \citep{heylShaviv, refractionDipoleEq}, it is generally believed that plasma effects should also become important deep within magnetar atmospheres \citep{laiHo03}. A particularly interesting alternative to the magnetar model describes AXP and SGR behaviour in terms of a quark star \citep{ouyed04, leahyOuyed09}. The environments around these objects are expected to be complex, and some models explain magnetar-like behaviour through accretion from fall-back material on to the surface of the object \citep{OLN07}. Interest in the quark nova scenario has grown considerably due in part to the \textit{NuSTAR} observations of the Cassiopeia A supernova remnant \citep{nuStarCasAObs} which detected a discrepancy in the distribution of titanium-44 and iron-56 abundances that can be explained by the quark nova scenario \citep{nuStarQuarkStars}. Considerations of more realistic dispersion relations coupled with GR may be relevant in the dynamic environment around an energetic, highly compact quark star.

\section{Conclusions}

Ray--tracing through the Schwarzschild space--time including plasma demonstrates light ray paths where refraction can dominate over gravity, scattering incoming rays. Figs. \ref{rayTraceh1} and \ref{rayTrace} show the effects for density power-law index $h=1$ and $h=3$, respectively. We have explored the orbits discussed in Section \ref{sec:orbits} with both the integral formalism (equation \ref{Delta}) and by solving the system of differential equations given by equations \ref{vel} and \ref{accel} directly using a fourth-- order Runge--Kutta method. The integral approach is valuable for analytical results like the pulse profiles given in Section \ref{sec:pulse}, while the RK4 scheme is suited to handling general cases when spherical symmetry is not assumed (for example, with the Kerr metric), and will be useful for more complicated plasma distributions.

In Section \ref{circularOrbits}, we found the radii of circular orbits for arbitrary $h$ and derived the turning points for trajectories that do not reach an observer at infinity. Depending on the details of the density distribution the plasma frequency can act as an effective photon mass \citep[$h=0$;][]{kulsrudLoeb}, an additional contribution to the angular momentum ($h=2$), or reproduce the potential of a massless scalar field perturbation on the Schwarzschild background with a judicious choice of scaling constant ($h=3$).

The interaction between a refractive medium and gravitational lensing has been considered at a variety of scales in the literature \citep{BB03, BKT09, mao14}. The effect of plasma on the formation of gravitational lensed images of background sources will be difficult to detect due to the small angular size of the images. Observing frequency--dependent magnification from transient microlensing events requires multi-waveband monitoring \citep{plasmaLensingKerr}. In this work we investigated another scenario in which dispersion and general relativistic effects may overlap: the pulse profiles generated by a compact object embedded in a continuous and smoothly varying refractive medium. We made use of the Schwarzschild metric to study the frequency--dependent effects, but a similar analysis can be done using the Reissner Nordstrom and Kerr \citep{plasmaLensingKerr} solutions. Both charge and rotation influence the space--time geometry and therefore would produce unique lensing effects \citep{RN_kerr_lensing}. This work shows that inclusion of even the simplest plasma effects generates achromatic behaviour in pulse profiles and novel trajectories for light rays near massive compact objects.

\section*{Acknowledgements}
I acknowledge and thank Samar Safi-Harb for numerous contributions to this work including discussion, proofreading and support through the Natural Sciences and Engineering Research Council of Canada (NSERC) Canada Research Chairs Program. I would also like to acknowledge the contributions of T. A. Osborn, Charlene Pawluck and Brian Corbett for proofreading and helpful suggestions for improving the text. Further thanks to Andrew Senchuk for many stimulating conversations during the development of this work. Finally, I would like to thank the anonymous referee whose constructive feedback improved the clarity and flow of the text.

\label{lastpage}

\end{document}